\documentclass[aps,prl,twocolumn,groupedaddress]{revtex4}
\usepackage[usenames,dvipsnames]{color}
\usepackage{graphicx}
\usepackage{subfigure}
\usepackage{verbatim}
\usepackage{epstopdf}

\begin{document}

\title{Network motifs come in sets: correlations in the randomization process}

\author{Reid Ginoza}
\email{reid.ginoza@gmail.com}
\affiliation{Bennington College, Bennington, VT 05201}

\author{Andrew Mugler}
\affiliation{Department of Physics, Columbia University, New York, NY 10027}

\date{\today}

\begin{abstract}
The identification of motifs---subgraphs that appear significantly more often in a particular network than in an ensemble of randomized networks---has become a ubiquitous method for uncovering potentially important subunits within networks drawn from a wide variety of fields.  We find that the most common algorithms used to generate the ensemble from the real network change subgraph counts in a highly correlated manner, so that one subgraph's status as a motif may not be independent from the statuses of the other subgraphs.  We demonstrate this effect for the problem of 3- and 4-node motif identification in the transcriptional regulatory networks of {\it E. coli} and {\it S. cerevisiae} in which randomized networks are generated via an edge-swapping algorithm (Milo et al., {\em Science} {\bf 298}:824, 2002).  We show that correlations among 3-node subgraphs are easily interpreted, and we present an information-theoretic tool that may be used to identify correlations among subgraphs of any size.
\end{abstract}

\maketitle

Identifying motifs has become a standard way to probe the functional significance of biological, technological, and sociological networks \cite{Wuchty:2003p254, Ma:2004p257, Sporns:2004p256, Berg:2004p258, YegerLotem:2004p259, Tsang:2007p255, Milo:2002p58}. A motif is commonly defined as a subgraph whose number of appearances in a particular network is significantly greater than its average number of appearances in an ensemble of networks generated under some null model \cite{Milo:2002p58}.  The typical null model prescribes an algorithm by which many randomized networks can be produced from the original network (see, e.g., Milo et al.\ \cite{Milo:2004p106} for a review and comparison of several such algorithms). While using an ensemble generated from the actual network often preserves features of the network that are desired for fair comparison (e.g.\ the degree distribution), this method may also induce unintended correlations in subgraph counts that ultimately influence the labeling of subgraphs as motifs.  The purpose of this note is to demonstrate and interpret such correlations in a simple case and describe how mutual information may be used to identify such correlations in general.

\begin{figure}
\centering
\includegraphics[scale=.54]{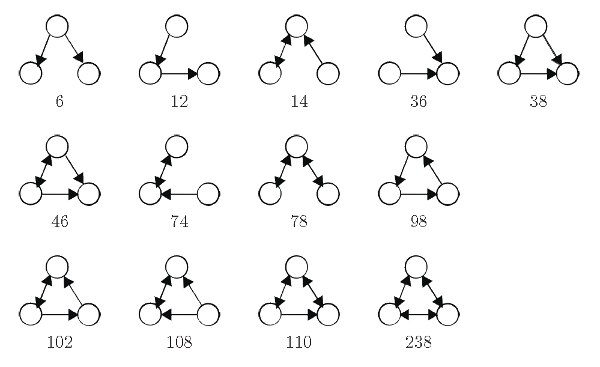}
\caption{All possible 3-node subgraphs, labeled as in Alon et al.'s ``motif dictionary'' \cite{AlonSite}.}
\label{dict}
\end{figure}

\section{Methods}

Following Milo et al.\ \cite{Milo:2002p58}, we perform 3- and 4-node motif detection on the transcriptional regulatory networks of {\it E. coli} (version $1.1$) and {\it S. cerevisiae}, using their freely available network data and software ({\tt mfinder} version $1.2$) \cite{AlonSite}.

Generation of randomized networks from the actual network is performed according to one of three null models: an edge-swapping algorithm, an edge-matching algorithm, and a Monte Carlo algorithm, all described in detail in \cite{Milo:2004p106}.  Because significance results are similar among models (cf.\ {\it Results} and \cite{Milo:2004p106}), emphasis in this note is placed on the edge-swapping algorithm, a Markov Chain procedure that repeatedly swaps the target nodes between pairs of edges.  $Z$-scores are computed from the mean and standard deviation of the count of a particular subgraph within an ensemble of at least 1,000 randomized networks \cite{Milo:2002p58}.

We quantify correlation between the counts of any two subgraphs over the course of a randomization process using mutual information \cite{Shannon:1949p69}.  Mutual information captures correlation between two random variables even when a relationship exists that is nonlinear (unlike, e.g., the correlation coefficient) or non-monotonic (unlike, e.g., Spearman's rho).  In this study, the counts $n_i$ and $n_j$ of the $i$th and $j$th subgraphs at each iteration of the edge-swapping process are used to increment a counts matrix from which the joint probability distribution $p(n_i,n_j)$ is obtained by normalization.  Mutual information $I_{ij}$ is computed as
\begin{equation}
\label{I}
I_{ij} = \sum_{n_i,n_j}p(n_i,n_j)\log_2\frac{p(n_i,n_j)}{p(n_i)p(n_j)},
\end{equation}
where the log is base 2 to give $I_{ij}$ in bits, and $p(n_i) = \sum_{n_j} p(n_i,n_j)$ and $p_{n_j} = \sum_{n_i} p(n_i,n_j)$.

Mutual information is bounded from below by $0$ (as seen in Eqn.\ \ref{I} when there is no correlation and the subgraph counts are independent of each other i.e.\ $p(n_i,n_j) = p(n_i)p(n_j)$) and bounded from above by the smaller of the two variables' entropies, where
\begin{equation}
\label{H}
H_i = -\sum_{n_i}p(n_i)\log_2 p(n_i)
\end{equation}
and the analogous expression with $i\rightarrow j$ are the entropies of the $i$th and $j$th subgraphs' counts respectively.  In order to obtain a statistic that can be compared across all subgraph pairs, we normalize by the average entropy, defining
\begin{equation}
\label{a}
a_{ij} = \frac{I_{ij}}{(H_i+H_j)/2}
\end{equation}
as our measure of correlation.  Note that $0 \le a_{ij} \le 1$, with $a_{ij}=0$ when $I_{ij}=0$ and $a_{ij}=1$ when $i=j$.  We find qualitatively similar results (cf.\ {\it Results}) when normalizing by the minimum, instead of the average, entropy.

\section{Results}

\subsection{An interpretable correlation}

\begin{figure}
\centering
\includegraphics[scale=.47]{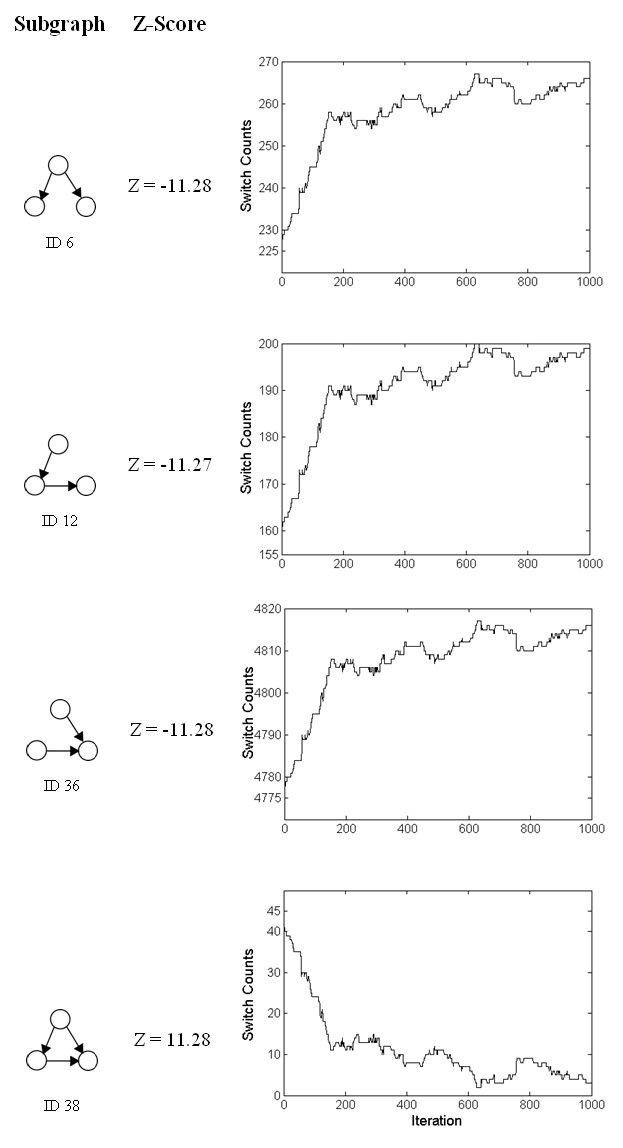}
\caption{The four 3-node subgraphs that appear in the {\it E. coli} network. $Z$-scores are calculated with respect to an ensemble of 1,000 randomized networks, generated via the edge-swapping algorithm \cite{Milo:2004p106}.  Plots  show the count of each subgraph during the generation of one randomized network.  Each iteration corresponds to one edge-swap.}
\label{table}
\end{figure}

\begin{figure}
\centering
\includegraphics[scale=.4]{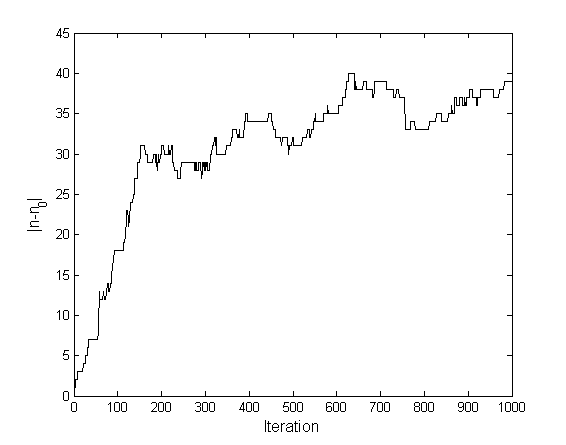}
\caption{Absolute value of the difference between count $n_0$ in the actual network and count $n$ at each iteration of the edge-swapping algorithm, for the subgraphs in Fig.\ \ref{table}.  Note that all four curves completely overlap.}
\label{overlap}
\end{figure}

\begin{figure}
\centering
\includegraphics [scale=.6] {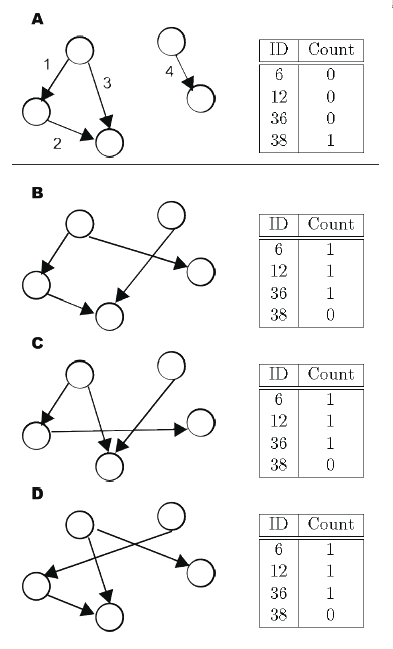}
\caption{Illustration of a correlation-producing effect.  Panel A shows subgraph 38 (the feed-forward loop) and an external edge, between which there are three possible edge swaps: a swap of edges 3 and 4 (panel B), a swap of edges 2 and 4 (panel C), and a swap of edges 1 and 4 (panel D).  In all three cases B, C, and D, subgraph 38 is reduced by one count, and subgraphs 6, 12, and 36 are each increased by one count.}
\label{toy}
\end{figure}

Only four 3-node subgraphs are present in the transcriptional network of {\it E. coli}, and a $Z$-score analysis of the type performed in \cite{Milo:2002p58} reveals a curious effect.  Specifically, with respect to ensembles generated via any of the edge-swapping, edge-matching, and Monte Carlo algorithms \cite{Milo:2004p106}, the $Z$-scores of three of the subgraphs (IDs 6, 12, and 36; cf.\ Fig.\ \ref{dict}) are either very close or equal to the negative of the $Z$-score of the fourth subgraph (the feed-forward loop, ID 38); see Fig.\ \ref{table}.  In fact, as shown in Fig.\ \ref{overlap}, the absolute value of the difference in counts within the actual network and counts within a sample randomized network at each iteration of the edge-swapping algorithm is the same among all four subgraphs for the first 1,000 iterations.  The interpretation is simple: as detailed in Fig.\ \ref{toy}, each time an edge of a feed-forward loop is swapped with an external edge, the feed-forward loop is destroyed and one of each of the other three subgraphs is created; using subgraph IDs we may denote this process as
\begin{equation}
\label{eff1}
38 \rightarrow 6, 12, 36.
\end{equation}
Since this process accounts for the overwhelming majority of the changes in count of the latter three subgraphs, there is extremely high correlation among the counts of all four subgraphs in each randomized network, and the magnitudes of their $Z$-scores are very close.

\subsection{An information-theoretic tool}

\begin{figure}
\centering
\includegraphics[scale=.36]{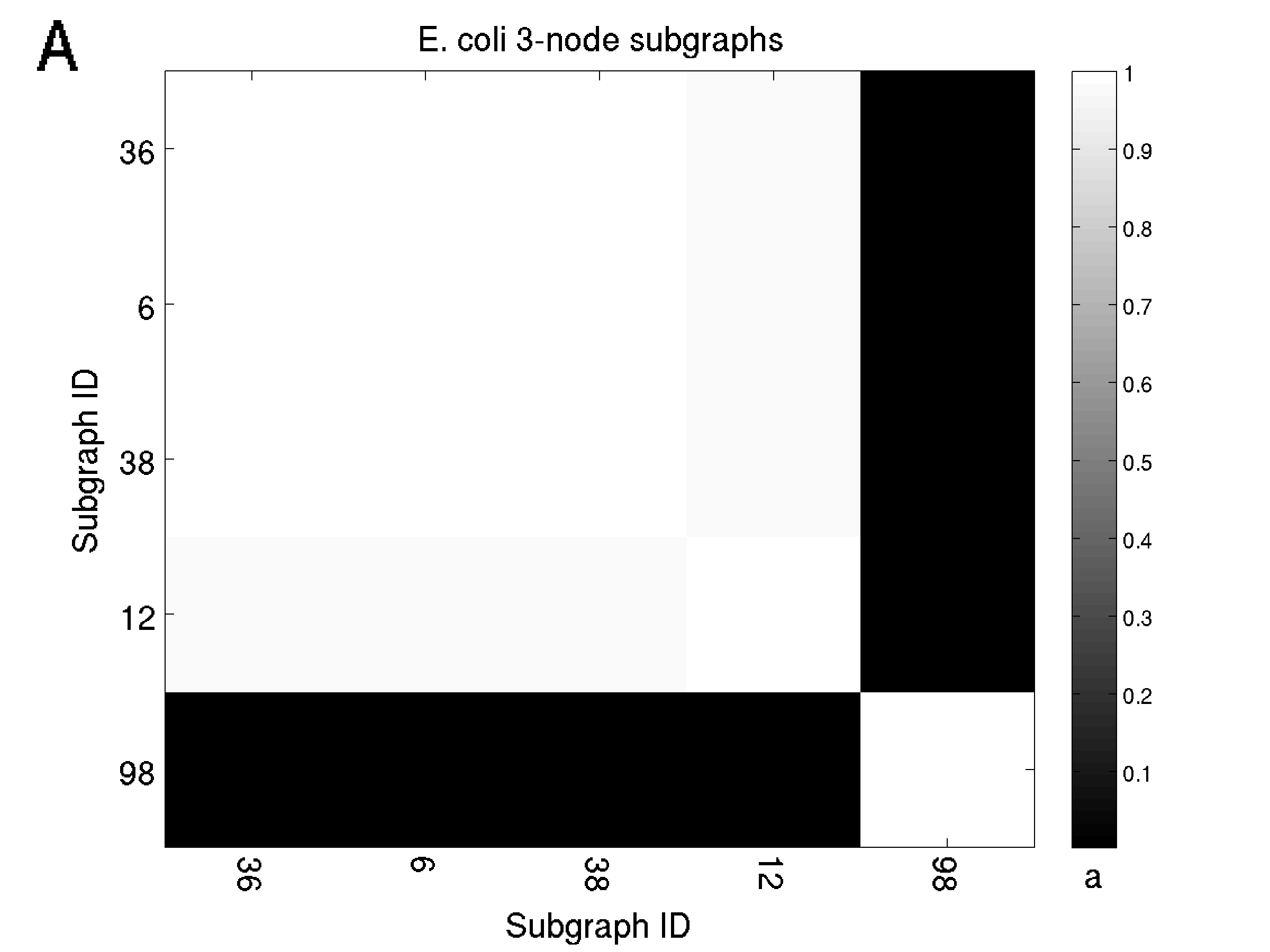}
\includegraphics[scale=.36]{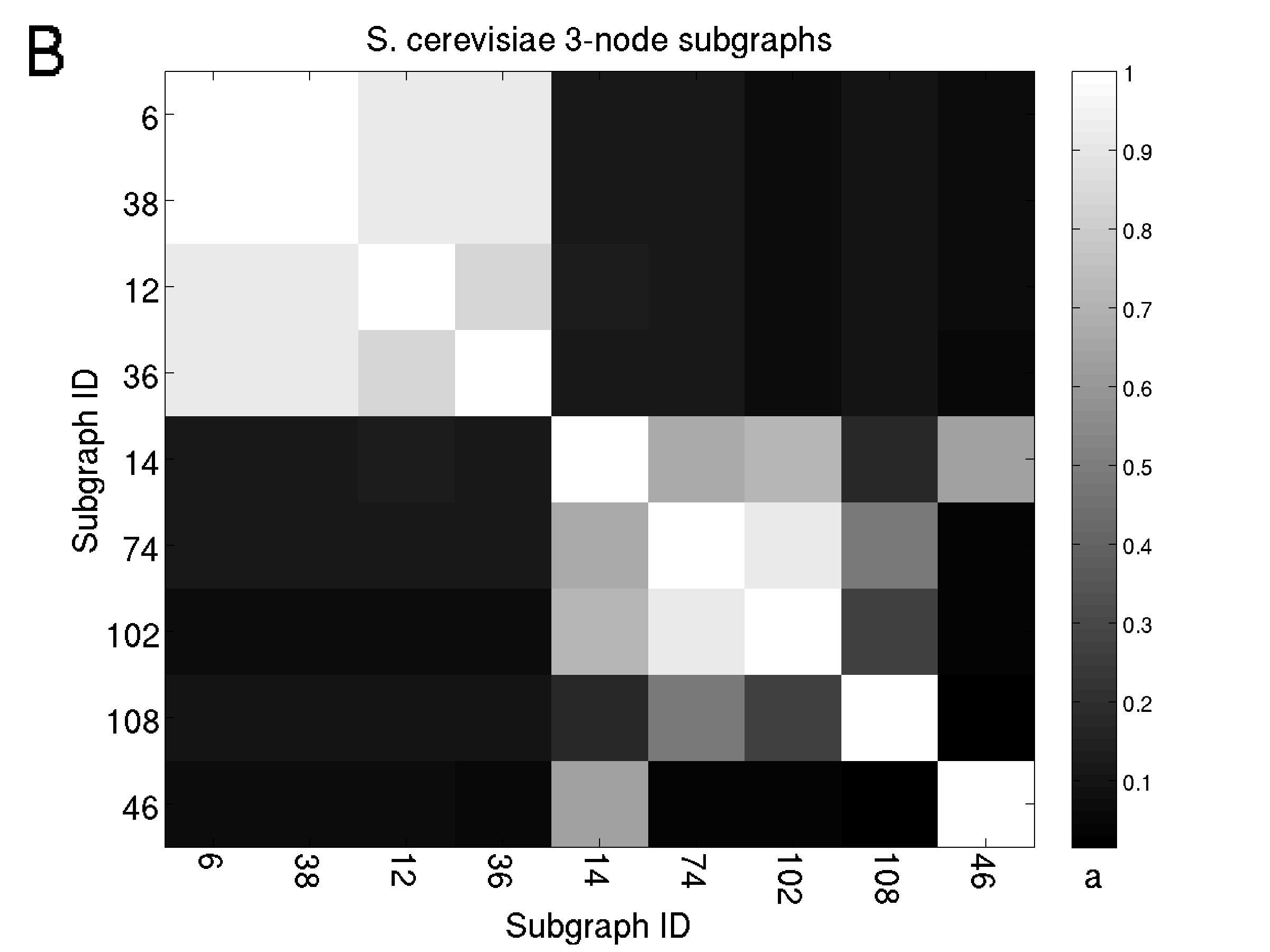}
\caption{Correlation measure $a_{ij}$ (cf.\ Eqn.\ \ref{a}) between all pairs of 3-node subgraphs $i$ and $j$ that appear during the randomization of a network via the edge-swapping algorithm \cite{Milo:2004p106} for the transcriptional networks of {\it E. coli} (A) and {\it S. cerevisiae} (B).  Subgraphs are labeled as in Fig.\ \ref{dict}.  The row and column order is determined by summing along either direction and sorting.}
\label{mi3}
\end{figure}

\begin{figure}
\centering
\includegraphics[scale=.36]{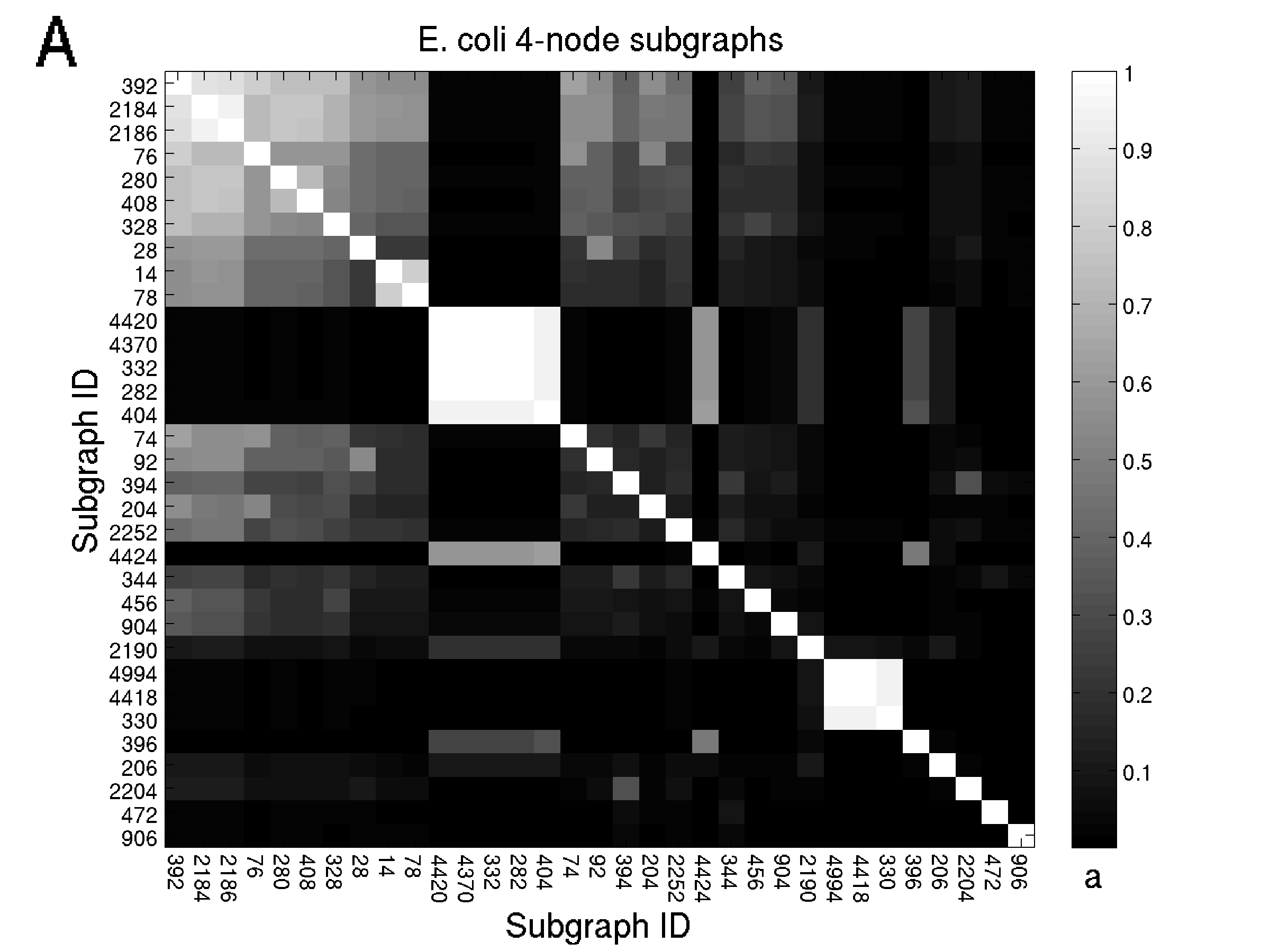}
\includegraphics[scale=.36]{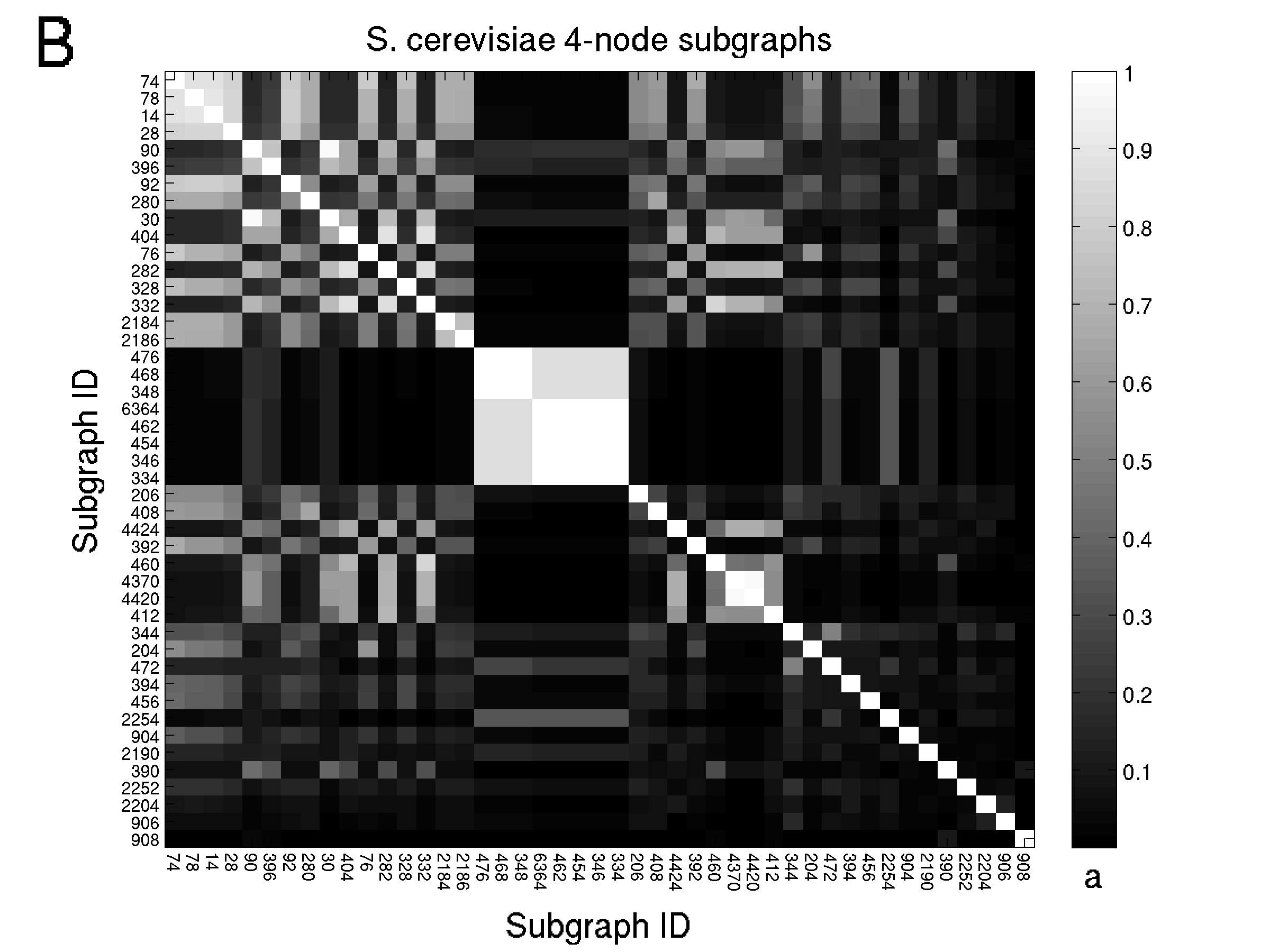}
\caption{Correlation measure $a_{ij}$ (cf.\ Eqn.\ \ref{a}) between all pairs of 4-node subgraphs $i$ and $j$ that appear during the randomization of a network via the edge-swapping algorithm \cite{Milo:2004p106} for the transcriptional networks of {\it E. coli} (A) and {\it S. cerevisiae} (B).  Subgraphs are labeled as in Alon et al.'s ``motif dictionary'' \cite{AlonSite}.  The row and column order is determined by summing along either direction and sorting.}
\label{mi4}
\end{figure}

To quantify and extend the detection of correlations such as that just described, we use a normalized mutual information measure, as detailed in {\it Methods}.  For the cases of 3- and 4-node subgraphs in both the {\it E. coli} and {\it S. cerevisiae} transcriptional networks, the measure $a_{ij}$ (cf.\ Eqn.\ \ref{a}) is computed between all pairs of subgraphs $i$ and $j$ that appear during the randomization of a network via the edge-swapping algorithm.  Figs.\ \ref{mi3}-\ref{mi4} show the matrices $a_{ij}$; the row and column order is determined by summing along either direction and sorting, which tends to group together sets of subgraphs with high pairwise correlations.

During randomization of the {\it E. coli} network, a set of four 3-node subgraphs (IDs 6, 12, 36, and 38; cf.\ Fig.\ \ref{dict}) are highly correlated, as shown by the bright 4-by-4 square in Fig.\ \ref{mi3}A.  The high correlation is simply the result of the effect described in the previous section, in which any of three swaps overwhelmingly converts a feed-forward loop (ID 38) into three other subgraphs (IDs 6, 12, and 36).  In fact the same set of high correlations in seen during the randomization of the {\it S. cerevisiae} network, as shown by the upper left 4-by-4 square in Fig.\ \ref{mi3}B.  There are additional correlated sets in {\it S. cerevisiae}: subgraphs 14, 74, and 102 are highly correlated as indicated by the bright 3-by-3 square involving these IDs in Fig.\ \ref{mi3}B, and subgraphs 74 and 108, as well as 14 and 46, are correlated as indicated by the relatively bright entries at these coordinate pairs in Fig.\ \ref{mi3}B.  Respectively, these correlations are due to the effects (in the notation of Eqn.\ \ref{eff1})
\begin{eqnarray}
\label{eff2}
102 &\rightarrow& 12, 14, 74,\\
108 &\rightarrow& 6, 74, 74,\\
46 &\rightarrow& 14, 14, 36,
\end{eqnarray}
of which one may convince oneself with the aid of Fig.\ \ref{dict}.  Note that although subgraphs 14, 102 and 108 participate in the highly correlated effects described here, none changes in number significantly enough upon randomization to be labeled a motif in the {\it S. cerevisiae} network (subgraphs 46 and 74 do not appear in the actual network, only during the course of the randomization).

Our analysis reveals correlations between counts of 4-node subgraphs as well.  As indicated by the bright blocks and off-diagonal elements in Fig.\ \ref{mi4}, several sets of subgraphs are highly correlated during the randomization of both the {\it E. coli} and {\it S. cerevisiae} networks.  Correlations are less easily interpreted in the 4-node case than in the 3-node case, but one must nonetheless remain aware of such artifacts of the randomization process when identifying subgraphs as motifs.  We note that the bi-fan (ID 204), the 4-node subgraph commonly identified as a motif in a variety of networks including both transcriptional networks studied here \cite{Milo:2002p58}, does not exhibit particularly high correlation with any other subgraph under our measure in either the {\it E. coli} or {\it S. cerevisiae} network.

We find results qualitatively similar to Figs.\ \ref{mi3}-\ref{mi4} when normalizing by the minimum, instead of the average, entropy in Eqn.\ \ref{a}.  The technique we describe here can be extended to the detection of subgraphs of any size.

\section{Discussion}

By quantifying correlations among subgraph counts during 3- and 4-node motif detection in the transcriptional networks of {\it E. coli} and {\it S. cerevisiae}, we reveal that motifs come in sets: the destruction of a subgraph during the randomization process can be highly correlated with the creation of one or more other subgraphs.  The correlations are easily understood in the 3-node case, and we present an information-theoretic tool to extract such correlations in general.  It has not escaped our attention that this observation serves as the basis for a more principled clustering of subgraphs based on correlations (e.g., by mixture-modeling in which the state of the subgraph count is a mixture of several states, with counts conditionally independent given the state).

The correlations among subgraphs are artifacts of the algorithm used to generate the ensemble of randomized networks; although we demonstrate their existence here in the context of only one randomization algorithm, the edge-swapping algorithm, they occur in other commonly used algorithms, as evidenced by mutually consistent effects on the $Z$-scores.  These findings do not necessarily invalidate the statuses of commonly identified motifs (it remains the case, for example, that there are significantly more feed-forward loops in the transcriptional network of {\it E. coli} than in a random network generated under most any commonly used null model); they do argue, however, that the limitations of the randomization scheme should be fully recognized during the motif finding process.

\acknowledgments
We thank Chris H.\ Wiggins for his helpful suggestions and guidance with this work.
AM was supported by NSF Grant No.\ DGE-0742450.

\bibliographystyle{apsrev}

\end{document}